\documentclass[runningheads]{llncs}
\usepackage[T1]{fontenc}
\usepackage{graphicx}
\usepackage{booktabs}    
\usepackage{threeparttable} 
\usepackage{amsmath}
\usepackage{epsfig}
\usepackage{csquotes}
\usepackage{enumitem}
\usepackage{tikz}
\usetikzlibrary{positioning}
\usepackage{caption}
\usetikzlibrary{calc}
\usepackage{graphicx}
\usepackage{subcaption}  
\begin{document}
\title{Haptics in Cognition: Disruptor or Enabler of Memory?}
%
%
\author{
Bibeg Limbu\inst{1}\orcidID{0000-0002-1269-6864} \and
Irene-Angelica Chounta\inst{1}\orcidID{0000-0001-9159-0664} }
\authorrunning{Limbu \& Chounta}
%
\institute{Department of Human-centered Computing and Cognitive Science, University of Duisburg-Essen, Germany \\\
\email{\{bibeg.limbu, irene-angelica.chounta\}@uni-due.de}}
\maketitle              
\begin{abstract}\
This exploratory pilot study investigates the impact of haptic perception — specifically tactile sensitivity (touch) and kinaesthetic intensity (movement)— on learning, operationalized as information retention (immediate recall) through handwriting. Participants (N=20) were randomly assigned to one of four experimental groups in a 2x2 factorial design, manipulating touch (via glove use) and movement (via increased writing pressure). Information retention was measured using an immediate recall test, while mental effort (reaction time in a secondary task) and perceived workload (NASA-TLX) were examined as mediating variables. Bayesian binomial regression revealed moderate evidence that increased writing pressure negatively influenced recall (85–88\% probability of negative effect), whereas glove use alone demonstrated no clear effect. Bayesian mediation analysis found no strong evidence that mental effort or perceived workload mediated these effects, as all 95\% credible intervals included zero, indicating substantial uncertainty. These findings suggest that increased Kinaesthetic demands may slightly impair immediate recall, independent of perceived workload or mental effort. Importantly, the manipulation of touch alone does not appear to influence information retention. The study contributes to understanding the nuanced relationship between embodied interactions and cognitive outcomes, with implications for designing sensor-based multimodal learning environments. 

\end{abstract}
\section{Introduction}\label{secIntro}


Haptic perception is typically understood as encompassing both tactile and kinaesthetic modalities \cite{loomis_lederman_1986}. The tactile modality, or simply touch, provides awareness of stimulation of the outer surface of the body, whereas the kinaesthetic sense, or simply movement, provides an awareness of static and dynamic body posture. Robert \cite{jutte_haptic_2008} draws attention to the ambiguous philosophical status of touch, a key aspect of Haptic perception alongside movement \cite{loomis_lederman_1986}. He notes that touch has occupied a paradoxical position, being alternately dismissed as the most negligible but also revered as the fundamental of the senses. Thomas Aquinas — a 13\textsuperscript{th}-century Italian philosopher and theologian — sought to reconcile Aristotle’s hierarchical senses in which sight often held primacy, into a coherent model where all sensory inputs — including touch — play a vital role in shaping the understanding of reality. He argued for the precedence of touch based on its optimal performance in the process of gathering knowledge, as touch roots perception in the entire bodily system. Aligning with Aquinas’s view, Geza Révész stressed that knowledge acquired through the sense of touch was more convincing and persuasive \cite{revesz_1944}. To form coherent mental models and/or cognitive representations through haptic senses, a vital process in encoding and storage of information in memory, and consequently learning, an agent must actively explore using both touch and movement \cite{kaas_neural_2008}. 

According to \cite{van2014four}, the acquisition of complex skills takes place in authentic real-world settings. Complex skills comprise a collection of cognitive and psychomotor skills with a collective goal, as is often the case in most real-life skills. For example, playing badminton often consists of knowing how to swing, move on the court, and having knowledge of the rules. Acquiring badminton skills requires playing badminton; that is, learning must be situated. Situated Cognition posits that learning is situated, goes hand in hand with doing, and is embedded in activities bound to social, cultural, and physical contexts \cite{greeno_situativity_1993}. Such learning, regardless of domain (for example, sports or mathematics), emerges through action \cite{Abrahamson_Lindgren_2014}, in which the human body acts as an actuator for the brain to act on the world and collects high-quality information through the haptic sense \cite{revesz_1944}. 

The body and the haptic senses play a vital role in learning and, by extension, to cognition as well. According to \cite{wilson_embodied_2013}, (embodied) cognition is not the body influencing the mind, but that the body and environment are part of the cognitive system itself. Therefore, cognition can be viewed as a dynamic, distributed process that depends on embodied interactions across the brain, body, and environment. \cite{wilson_embodied_2013} posits cognition as situated and task-specific, arising from solving concrete, real-world problems (see \textquote{4E cognition} by \cite{carney_thinking_2020}). At the same time, embodied cognition has been criticized for its lack of formalization and theoretical rigor, rendering many of its claims vague and unfalsifiable \cite{carney_thinking_2020,hommel_theory_2015}. For example, it is unclear how bodily states translate into cognitive processes. Embodiment theories often reject classical cognitivist frameworks—such as internal representations and symbolic processing—without offering sufficient empirical justification \cite{carney_thinking_2020}. \cite{hommel_theory_2015} instead advocates combining embodied and classical cognitivist models, rather than treating them as mutually exclusive. In line with these perspectives, we explore the following research questions:

\begin{itemize} [labelindent=2em, leftmargin=3em]
\item[\textbf{RQ1:}] How does the sensitivity of tactile perception affect Information Retention? 
\item[\textbf{RQ2:}] How does the intensity of motor activity affect Information Retention? 
\end{itemize}

As stated earlier, Haptic senses consist of tactile and Kinesthetic modalities \cite{loomis_lederman_1986}. We explore how the two dimensions of haptic sense impact cognition, specifically information processing in memory. First, we examine the impact of the sensitivity of tactile perception (touch) on mental effort and information retention (\textbf{RQ1}). As Tactile and Kinesthetic modality (movement) are tightly intertwined during an embodied learning activity, we secondly examine the impact of the intensity of motor activities during movement on mental effort and information retention (\textbf{RQ2}). Our work aims to contribute to our understanding of the cognitive mechanisms by which the bodily state impacts cognition and learning. 

The rest of the paper is structured as follows: in Section \ref{secBackground}, we provide the background necessary for the study, followed by the method of study (see Section \ref{secMethod}). Then, we present the results of the data analysis in Section \ref{secResult}, followed by a discussion on the theoretical and practical implications and limitations of this work (Section \ref{secDiscuss}).

\section{Related Work}\label{secBackground}

Information retention in memory is affected by the following four cognitive processes: encoding, consolidation, storage, and retrieval \cite{brem_learning_2013}. Encoding is a vital process for the acquisition of new information, which transforms new information into mental representations to be stored in memory for later recall. Embodied cognition suggests that sensorimotor experiences, such as with haptic senses, contribute significantly to the formation of cognitive representations during the encoding process (see Grounded Cognition by \cite{barsalou2008grounded}). Integration of multiple senses has been found to improve task performance in complex skills, particularly in high-load conditions \cite{siqueira_rodrigues_comparing_2024,marucci_impact_2021}. Successful recall of domain-specific information from memory is a requisite for demonstrating task performance in complex skills \cite{sawyer_zig_2013,tricot_domain-specific_2014}. Moreover, practice in a familiar context in which the learner was situated or embedded during the learning phase can significantly improve recall of information from memory \cite{horz_situated_2012}. 

Although the use of multiple senses may enhance learning in specific scenarios, Vermeulen et al. \cite{vermeulen_sensory_2008}, in contrast, also found that sensory overload often leads to higher mental effort. The encoding process, much like the other three cognitive processes, exerts mental effort and impacts the assimilation, retention, and retrieval of information in/from long-term memory (see Cognitive Load Theory \cite{sweller_cognitive_2011}). Mental effort can be conceptualised as how hard a person tries to actively process incoming information in working memory \cite{kirschner_mental_2012}. Working memory holds a limited amount of information during encoding \cite{baddeley_human_1990}, and requiring a learner to hold more information can lead to cognitive overload, which is detrimental for learning. \cite{baddeley_human_1990}'s model of working memory, as depicted in his 1990 book \textquote{Human Memory: Theory and Practice}, ignored the role of haptic senses, only accounting for the verbal \& auditory information, and the visual \& spatial information. \cite{baddeley2003working} later acknowledged the scarcity of tactile-working-memory research and emphasized the need for multimodal integration, which implies including haptic modalities in the working memory model. Other relevant works have alluded to haptic working memory being a limited-capacity system, similar to Baddeley's working memory model \cite{lerch_exploring_2016,sebastian_working_2008}. \cite{morimoto_nature_2020,sebastian_working_2008} found that performing two distinct tasks within the same sensory modality leads to significant interference due to limited modality-specific capacity, while performing tasks across different modalities (e.g., visual and haptic) also causes interference, though typically to a lesser extent. Their findings support the idea of a partially shared, capacity-limited working memory system with modality-specific buffers, similar to the visual and auditory buffers in working memory (see Dual-coding theory \cite{clark_dual_1991}). This suggests that high bodily involvement or physical activity—particularly when not directly related to the learning task—may impede learning by contributing to cognitive overload \cite{skulmowski_embodied_2018}. In embodied learning, as is often the case in situated learning and authentic practice, multiple senses and motor activities are involved. How the inclusion of additional haptic sense modalities during learning impacts mental effort and information retention remains unclear. To further explore this gap, in this study, we explored the association between haptic senses and memory through the context of handwriting.

\subsection{Handwriting and Memory}
Handwriting remains a prevalent motor activity in educational contexts and has proven beneficial for learning \cite{skar_handwriting_2022,ray_relationship_2022}. This highlights the importance of internalizing competent handwriting fluency and is a prerequisite for all further handwriting-related learning benefits. Several studies \cite{mueller_pen_2014,wrigley_avoiding_2019,flanigan_typed_2024} found that students who took notes on laptops performed worse on conceptual questions than students who took handwritten notes. Similarly, other related research demonstrated that taking notes by hand -- instead of typing -- can lead to improved memory \cite{smoker_comparing_2009,bouriga_is_2021,van_der_weel_handwriting_2024}. This may partly explain why handwritten note-taking continues to be widely practiced even within the academic community \cite{yannic2025our}. The authors argued that laptop note-takers tend to transcribe lectures verbatim, which is detrimental to learning. In contrast, handwriting necessitates processing information and reframing it in their own words, as handwriting is slower than typing, and learners need to condense the incoming information. This additional mental effort requires learners to cognitively engage with the material, leading to improved retention of information in memory.

Learning strategies referred to as \textquote{Desirable Difficulties} \cite{bjork_making_2011} often use handwriting to slow down the learner, forcing learners to expend more mental effort, leading to deeper levels of processing, and consequently better memory \cite{craik_depth_1975}. However, higher-level cognitive processes involved in handwriting may also lead to cognitive overload, and instead inhibit the storage of information \cite{peverly_importance_2006}. Nonetheless, \cite{ose_askvik_importance_2020} found that handwriting induces theta-range synchronized activity in brain areas linked to memory and learning, which supports encoding of new information. Furthermore, handwriting resulted in increased activation in brain regions associated with language processing, working memory, and executive functions during handwriting tasks \cite{van_der_weel_handwriting_2024}. \cite{longcamp_influence_2005,smoker_comparing_2009} argued that the brain receives multiple sensory stimuli during writing (visual, motor, and kinaesthetic), which typing does not do in the same manner. Typing may limit the embodied reinforcement needed for robust cognitive-symbolic representations during encoding, particularly in young learners \cite{mangen_pen_2016}. 



\subsection{Sensors in Handwriting}

Sensors and actuators enhance authentic practice within situated learning environments by monitoring learners' multisensory experiences and, when necessary, engaging their perception through actuator-driven interventions \cite{specht2019sensors,limbu2018using,matsanusi2025virtual}. Multimodal Immersive Learning Environments (MILEs), using sensors and actuators, embed the learner in (simulated) realistic contexts for complex skills training \cite{dimitri2022multimodal,Limbu2019Conceptual,lee2023role}. For example, \cite{schneider_beyond_2019,limbu2023Hololearn,Fernando_2025_onyour} used virtual reality to embed the user in authentic contexts by simulating the contextual surroundings. MILEs are projected to play a key role in facilitating embodied learning and facilitating inquiry into (embodied) cognition theories \cite{giannakos_role_2023}. 

In the context of handwriting, digital stylus/pen with tablets are increasingly used to train, analyse, and inquire into the phenomenon of handwriting learning \cite{dikken2022deliberate,bonneton-botte_teaching_2023}. \cite{danna_basic_2015} propose various forms of supplementary stimuli, in addition to the naturally present primary stimuli, which can be introduced during handwriting with the help of MILEs \cite{danna_basic_2015,kiefer_writing_2016}. For example, \cite{loup-escande_contributions_2017,limbu2019canyou} have used color gradients in their applications to provide supplementary visual information about pen pressure while writing. As previously established, handwriting benefits memory by activating multiple sensory pathways \cite{longcamp_influence_2005,smoker_comparing_2009}, the potential benefit of providing supplementary stimuli encourages MILEs designers to make use of haptic senses. However, it is essential to exercise caution, as \cite{loup-escande_contributions_2017} found that the supplementary information led to an increase in mental effort. In contrast, \cite{limbu2019canyou,limbu2025mighty} found no significant difference in mental effort when supplementary stimuli was provided compared to not providing them. It should be noted that \cite{limbu2019canyou,limbu2025mighty} had limited participants to make definite conclusions. \cite{yoshida_tactile_2015} postulate a smaller memory capacity for haptic senses than for visual senses. A complex implementation of supplementary haptic stimuli can more easily create cognitive overload. While the advantages of handwriting on paper have been well studied, there is a lack of understanding of how those benefits will transfer to digital devices \cite{kiefer_writing_2016}. Consequently, sensors and MILEs are becoming indispensable for the study of embodied cognition theories, but the design of MILEs requires a nuanced understanding of the role of haptic senses and multisensory learning principles, to which this study seeks to contribute.

\section{Methods}\label{secMethod}
To investigate our research questions, we conducted a formative study in which we conceptualized haptic sense along two key dimensions: (a) the acuity of tactile perception (touch), and (b) the intensity of kinaesthetic engagement (movement). We compared learning performance across four groups—each defined by a unique combination of high or low levels of these two dimensions—using an immediate post-test. Quantitative data on participants’ mental effort and perceived workload were also collected and analyzed as potential mediating variables.

\subsection{Participants}
Twenty (20) right-handed, German native-speaking university students participated in the study. We invited only right-handed students to control the variation that may arise from the dominant hand. Twelve (12) participants were female, and eight (8) were male. The mean age of the female participants was $M_\text{female} = 21.6$ (SD = 1.83), and the mean age of the male participants was $M_\text{male} = 23.1$  (SD = 2.23).

\subsection{Experimental Design}\label{sec:ExpDesign}
This study investigated the effects of haptic senses on memory recall across two dimensions—tactile sensitivity and proprioceptive intensity—using a between-subjects 2x2 factorial design (see Table \ref{tab:ExpDesign}). The two treatment conditions were: (1) proprioceptive intensity, induced by increased handwriting pressure (P), and (2) reduced tactile sensitivity, achieved through the use of a glove (G). Participants in Group 1 applied additional handwriting pressure and wore a glove (P+G). Group 2 participants applied additional pressure without wearing a glove (P+¬G), while those in Group 3 wore a glove but did not apply additional pressure (¬P+G). The control group (Group 4) neither applied additional pressure nor wore a glove (¬P+¬G). Participants assigned to conditions requiring increased pressure received auditory feedback cues to maintain the required pressure level during the task. This feedback cue was acoustically distinct from the auditory stimuli used in the secondary task to avoid confusion (see Section \ref{secMentalEffort}).

\begin{table}[h!]
    \centering
    \caption{Experimental groups based on two factors: proprioceptive intensity (P) and tactile sensitivity (G) as seen in \cite{limbu2025mighty}}
    \label{tab:ExpDesign}
    \begin{tabular}{@{}l@{\hskip 12pt}c@{\hskip 12pt}c@{}}
        \toprule
        \textbf{Group} & \textbf{Condition} & \textbf{N} \\
        \midrule
        1 & P + G       & 5 \\
        2 & P + ¬G      & 6 \\
        3 & ¬P + G      & 5 \\
        4 & ¬P + ¬G     & 4 \\
        \bottomrule
    \end{tabular}
\end{table}

\subsection{Procedure}
Upon arrival, participants were briefed on the study's objectives and the experimental task procedures. Participants were informed of their rights, including the right to withdraw at any time, and were asked to provide written informed consent. The participants were assigned a unique identifier code, which was used to anonymize participants' data. The study received ethical approval from the ethics committee of the Department of Human-centered Computing and Cognitive Science, University of Duisburg-Essen (Ethics Committee Opinion ID: \textit{2407CMLB3526}).

Before the study began, the participants familiarised themselves with the apparatus. After that, they were randomly assigned to one of the study groups (Section \ref{sec:ExpDesign}). Participants were subsequently reminded of the primary task (copying the text displayed on the prompter) and the secondary task (responding to auditory stimuli by pressing a key on the keyboard). The text for the experimental task that the participants copied was displayed on the prompter sentence by sentence. Based on the treatment conditions/group allocation, participants were assigned to either wear a glove, write with extra pressure, do both, or write normally. Following the experimental task, participants responded to the questionnaires assessing their recall and perceived workload. The order of test administration was randomized to mitigate potential ordering effects, such as the influence of the recall task on workload ratings or vice versa. The entire procedure took approximately 30-45 minutes.

\subsection{Apparatus}

\begin{figure}[!h]
  \centering
   {\epsfig{file = 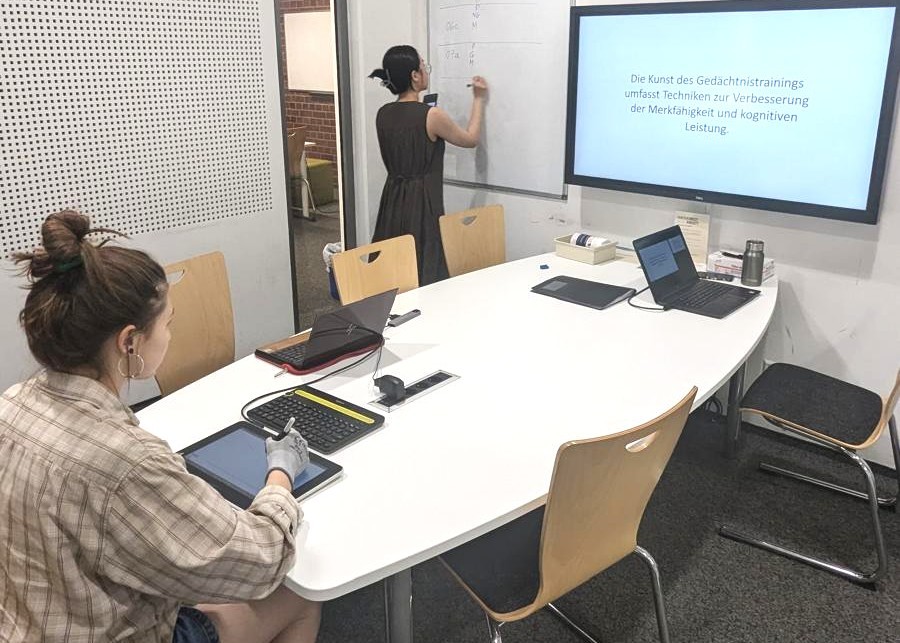, width = 10cm}}
  \caption{Experimental setup with a teleprompter displaying text for participants to copy using a stylus}
  \label{fig:ExpSetup}
\end{figure}

Participants used a WACOM One™ tablet and stylus to copy text from the teleprompter (Figure \ref{fig:ExpSetup}). The tablet was connected to a laptop that ran a custom application to log data. The application also reminded the participants in the \textquote{P} treatment group to exert more pressure when needed. The participants were instructed to press the spacebar key (secondary task) on the wireless keyboard placed next to them as quickly as possible in response to the auditory signal (Section \ref{secMentalEffort}). A gardening glove selected to exert minimal impact on the pen grip was used to reduce the tactile sensitivity for the groups with the \textquote{G} treatment.

\subsection{Materials and Measures}
\subsubsection{Mental Effort}\label{secMentalEffort}
In this study, \textit{Mental effort} was measured using the dual-task method. Mental effort is the cognitive capacity that is allocated to accommodate the demands imposed by the task \cite{kirschner_mental_2012}. The dual-task paradigm is a behavioural method used to assess mental effort over time by providing an additional secondary task (such as, reacting to auditory stimuli \cite{limbu2019canyou,limbu2025mighty}) to measure decay in secondary task performance (or primary) \cite{esmaeili2021currentview}. As a secondary task, participants reacted to the auditory stimuli as fast as possible by pressing the spacebar key on a wireless keyboard. The reaction time was logged in milliseconds.

\subsubsection{Perceived Workload}\label{sec:PerceivedWorkload}
\textit{Perceived workload} is the physical and mental effort invested and was measured using the NASA-TLX instrument \cite{hart1986nasa}. The NASA-TLX is a subjective instrument that evaluates perceived workload across six dimensions: \textit{mental demand, physical demand, temporal demand, performance, effort, and frustration}. Participants scored each dimension on a scale from 0 to 100, reflecting the perceived intensity of each factor. A pairwise comparison between the 15 pairs from 6 dimensions is then performed, in which the participant selects the most influential dimension from the two. Based on this, the individual weights for each dimension were calculated and then used to calculate the overall perceived workload.

\subsubsection{Information Retention}
Retention of learned information is defined as the storage of information in long-term memory such that it can be readily retrieved or recalled, for example, in response to standard prompts, and not merely in response to experiential cues \cite{bennett2012retention}. \textit{Information Retention} was measured using a multiple-choice knowledge test and was administered immediately after the experiment to assess the participant's ability to recall the information presented during the experiment. The knowledge test consisted of 10 questions related to the content presented during the experiment. For each question, there were four possible choices with only one correct answer. A control question was included to verify that responses were not given at random. Furthermore, participants were also instructed to skip the question if they did not know the answer rather than guessing.  

\section{Results}\label{secResult}
The data collected as part of this study is publicly accessible \cite{limbu_2025_data}. The R script used for the analysis of data can be found here\footnote{https://github.com/bibeglimbu/Haptics\_HWCompare\_Bayes.git}. We adopted the BARG (Bayesian Analysis Reporting Guidelines) for reporting our results \cite{kruschke_bayesian_2021}. We opted for Bayesian analysis as it is more robust to smaller sample sizes and does not depend on the P-values. 

\begin{table}[h!]
    \centering
    \caption{Mean and Standard Deviation for Immediate Recall,
Perceived Workload, and Mental Effort}
    \label{tab:ExpDesign}
    \begin{tabular}{@{}l@{\hskip 12pt}c@{\hskip 12pt}c@{\hskip 12pt}c@{}}
        \toprule
        \textbf{Group} & \textbf{Information Retention} & \textbf{Perceived Workload} & \textbf{Mental Effort} (sec)\\
         & (0-10) & (0-100) & (2.74-14.88) \\ 
        \midrule
         1 & \textbf{7.20}, \textit{1.64} & \textbf{57.2}, \textit{21.0} & \textbf{5.42}, \textit{1.18}\\ 
    \hline
    2  & \textbf{7.17}, \textit{1.72} & \textbf{46.8}, \textit{16.2} & \textbf{4.55}, \textit{1.12}\\ 
    \hline
    3 & \textbf{8.40}, \textit{1.52}& \textbf{40.2}, \textit{14.2} & \textbf{8.77}, \textit{4.07}\\ 
    \hline
    4 & \textbf{8.25}, \textit{1.71} & \textbf{52.0}, \textit{13.6} & \textbf{4.77}, \textit{2.09}\\ 
        \bottomrule
    \end{tabular}
\end{table}

\begin{figure}[!h]
  \centering
  
  \begin{subfigure}[b]{0.45\textwidth}
    \includegraphics[width=\linewidth, height=4.3cm]{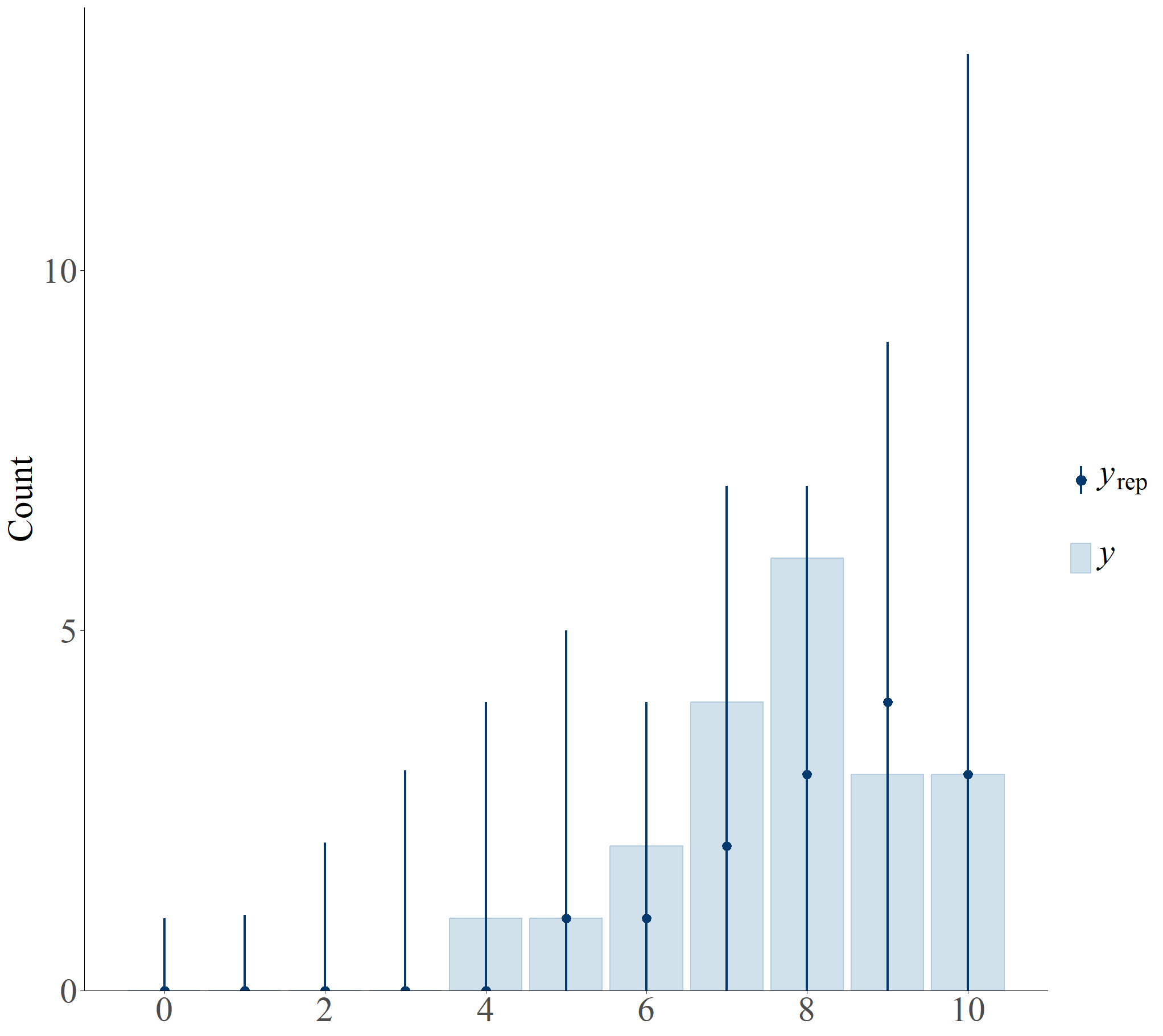}
    \caption{Binomial Prior for IR}
    \label{fig:sub1}
  \end{subfigure}
  \hfill
  \begin{subfigure}[b]{0.45\textwidth}
    \includegraphics[width=\linewidth, height=4.3cm]{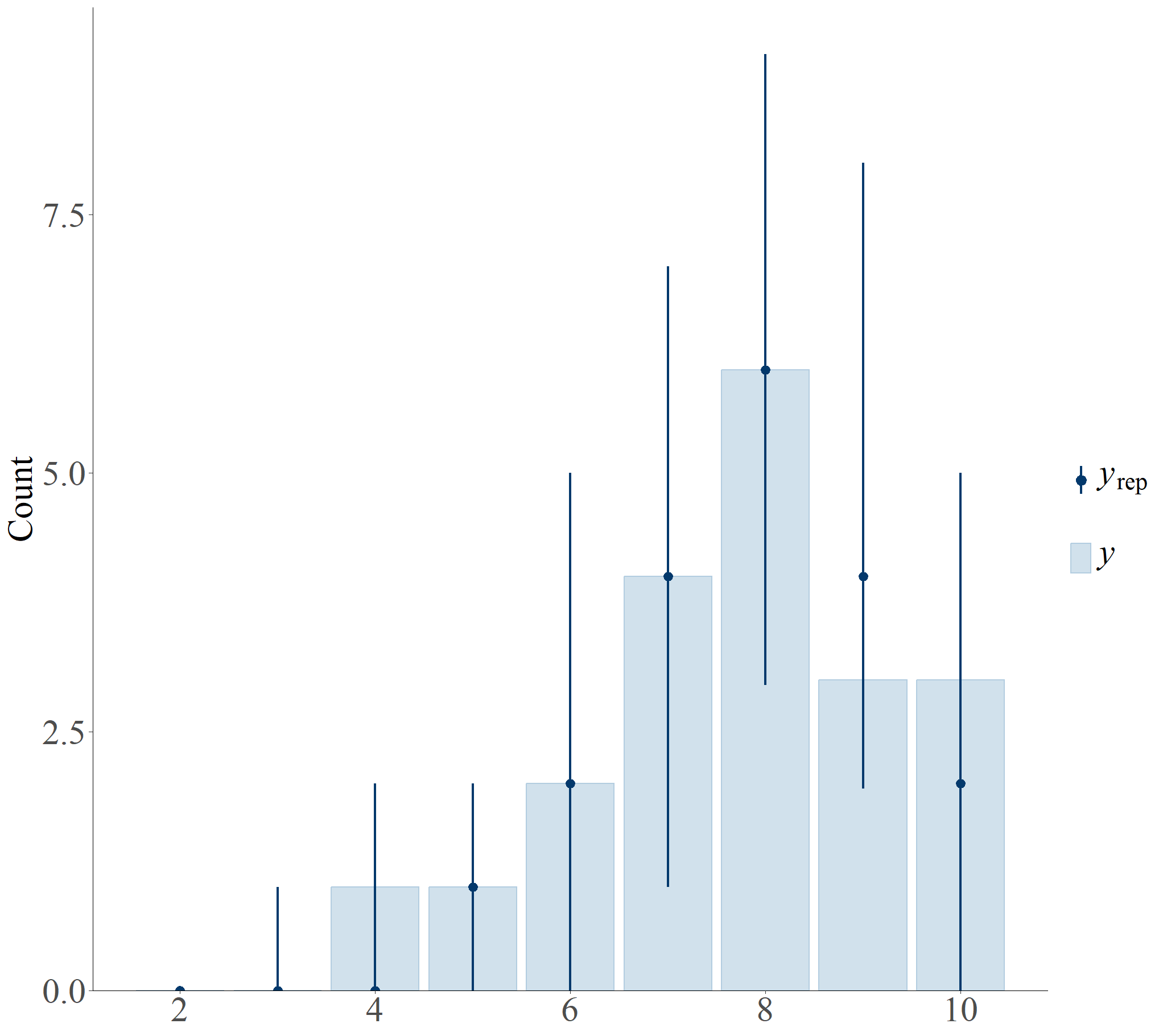}
    \caption{Binomial Posterior for IR}
    \label{fig:sub2}
  \end{subfigure}
  
\vspace{0.5cm}

  \begin{subfigure}[b]{0.45\textwidth}
    \includegraphics[width=\linewidth, height=4.3cm]{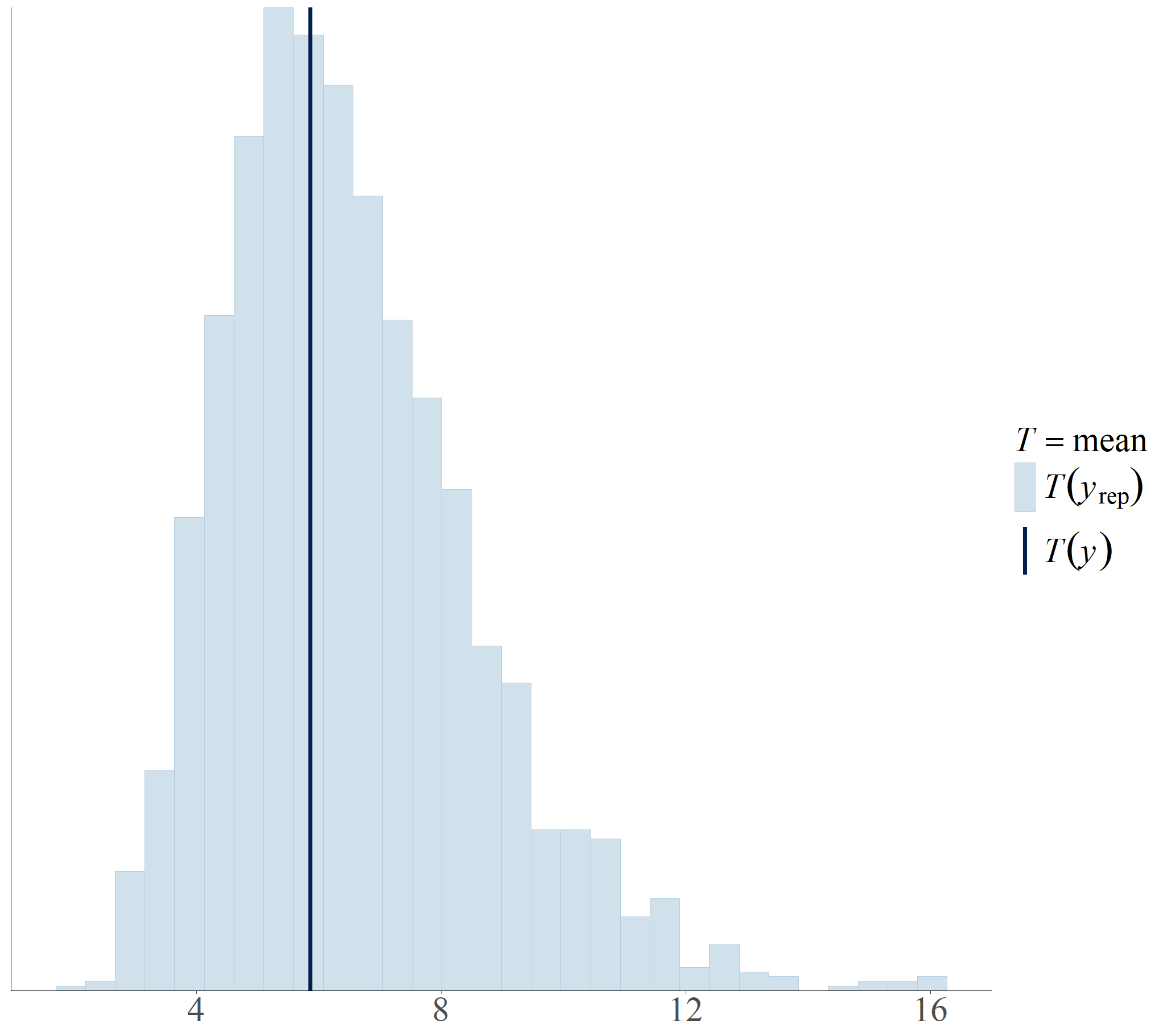}
    \caption{Lognormal prior for ME}
    \label{fig:sub3}
  \end{subfigure}
\hfill
  \begin{subfigure}[b]{0.45\textwidth}
    \includegraphics[width=\linewidth, height=4.3cm]{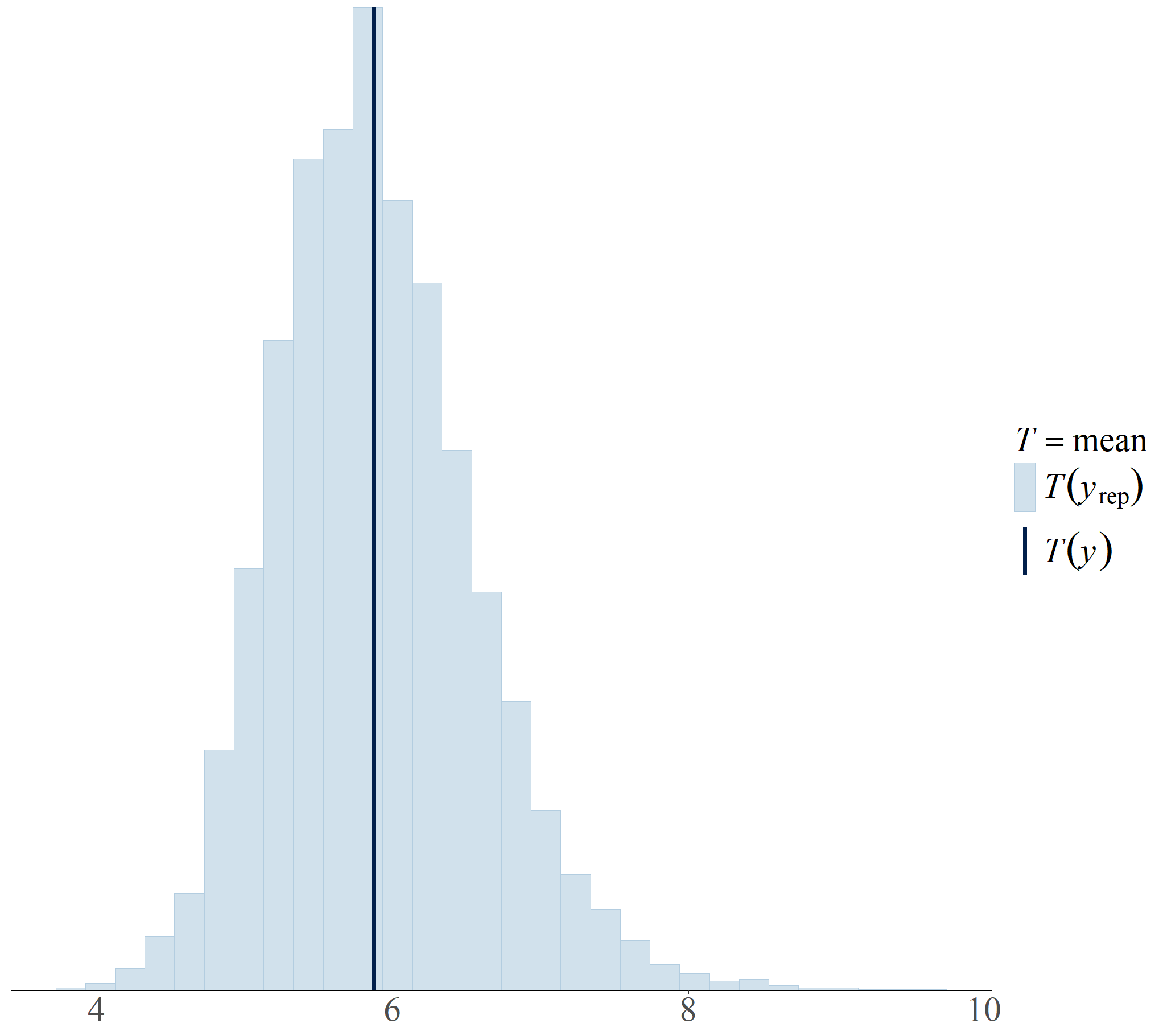}
    \caption{Lognormal Posterior for ME}
    \label{fig:sub4}
  \end{subfigure}
  
  \vspace{0.5cm}
  
  \begin{subfigure}[b]{0.45\textwidth}
    \includegraphics[width=\linewidth, height=4.3cm]{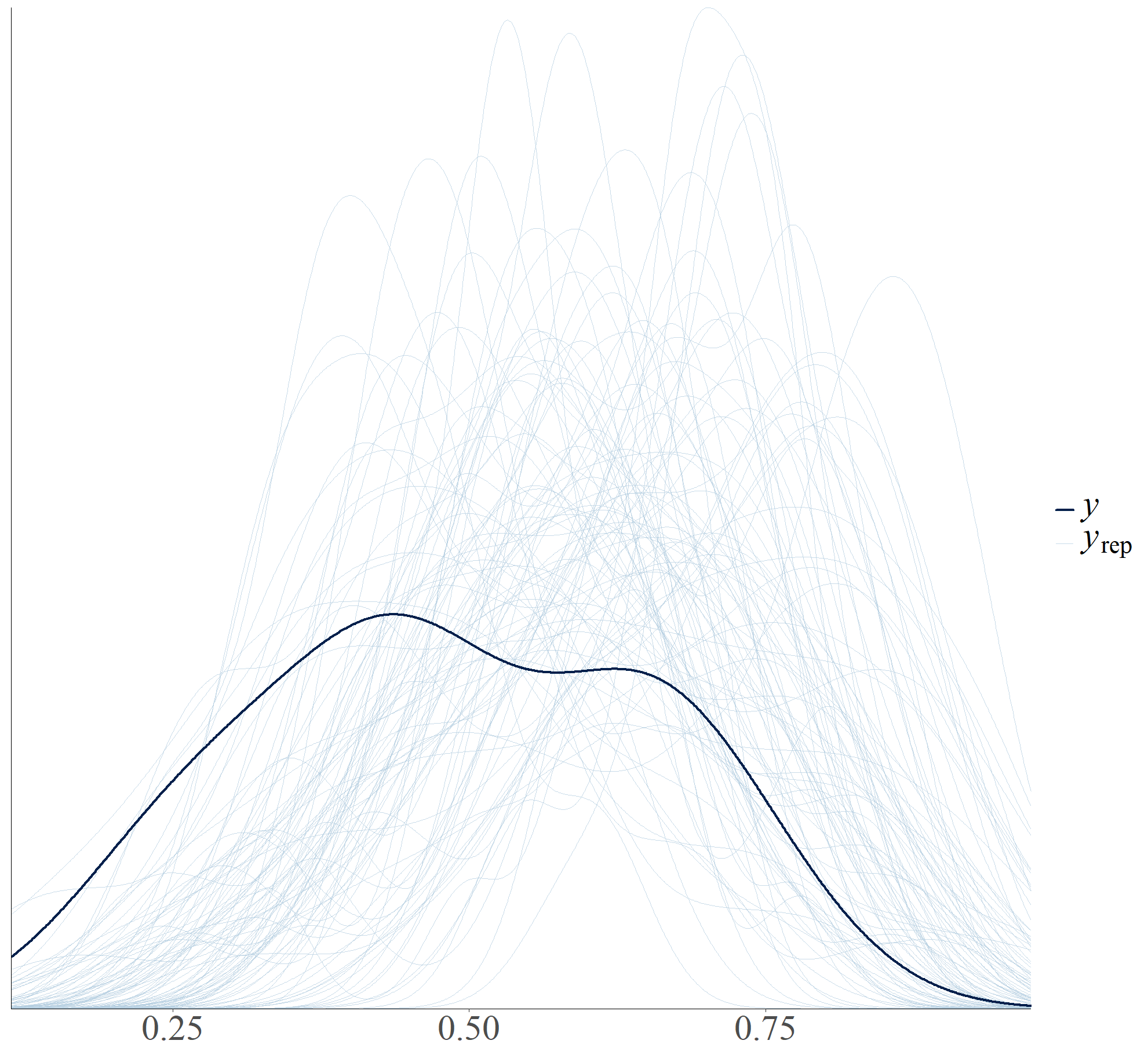}
    \caption{Beta Prior for PW}
    \label{fig:sub5}
  \end{subfigure}
  \hfill  
  \begin{subfigure}[b]{0.45\textwidth}
    \includegraphics[width=\linewidth, height=4.3cm]{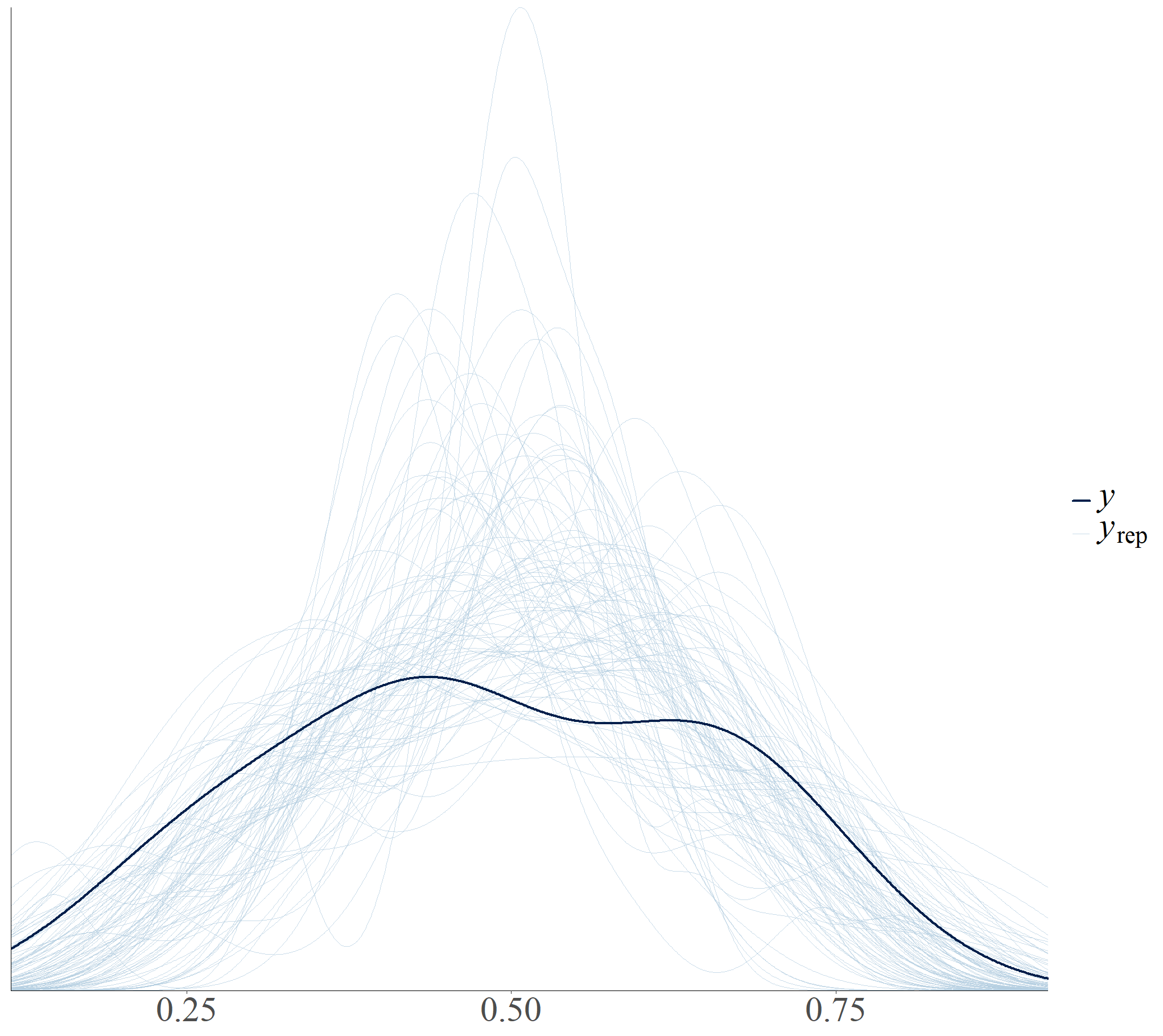}
    \caption{Beta Posterior for PW}
    \label{fig:sub6}
  \end{subfigure}

  \caption{Prior and Posterior distributions of Immediate Recall (IR), Mental Effort (ME), and Perceived Workload (PW) based on binomial, lognormal, and beta likelihoods, respectively}
  \label{fig:PriorPostDistribution}
\end{figure}

\subsection{Information Retention} \label{subsecInfoRetention}
We fitted a Bayesian binomial regression model with a logit link function using the brms package\footnote{https://cran.r-project.org/web/packages/brms/index.html} in R to examine group performance differences in Information Retention. Information Retention was measured as a post-test using a 10-item multiple-choice test. In the model, we initially included weakly informative priors: $normal(0, 1)$ for group-level effects and $normal(0, 5)$ for the intercept (control group/Group 4). Prior and posterior predictive checks were conducted to evaluate the plausibility of the initial priors (Figure \ref{fig:PriorPostDistribution}). We placed the intercept prior to $normal(1.5, 1)$, reflecting the belief that most participants would perform moderately well ($\sim80$\% correct) with moderate uncertainty.

We expected the group-level effects to be small, with most groups performing sufficiently well, as the text used for the study was not complicated and the only immediate recall was measured (Section \ref{secMethod}). As previously discussed in Section \ref{secBackground}, there is a mixed consensus on the true effect of handwriting on memory. However, there is a positive trend for the benefits of handwriting indicated by seminal papers such as \cite{mueller_pen_2014} and, more recently, evidence from cognitive neuroscience studies \cite{van_der_weel_handwriting_2024}. Therefore, the new prior for the intercept is fitting, based on evidence and theoretical assumptions, as all groups used handwriting. In contrast, we have very little to no information on the effect of the groups. The updated prior yielded predictions that aligned well with plausible pre-data expectations and passed the prior predictive check. Four Markov Chain Monte Carlo (MCMC) chains were run with 4000 iterations each (2000 warm-up), yielding 8000 post-warmup samples. All $\hat{R}$ values were 1.00, indicating good convergence. After examining posterior means and 95\% credible intervals (CI) for each group coefficient, we quantified directional evidence by computing the posterior probability that each effect was negative and the corresponding Bayes factors.

\begin{figure}[!h]
  \centering
   {\epsfig{file = 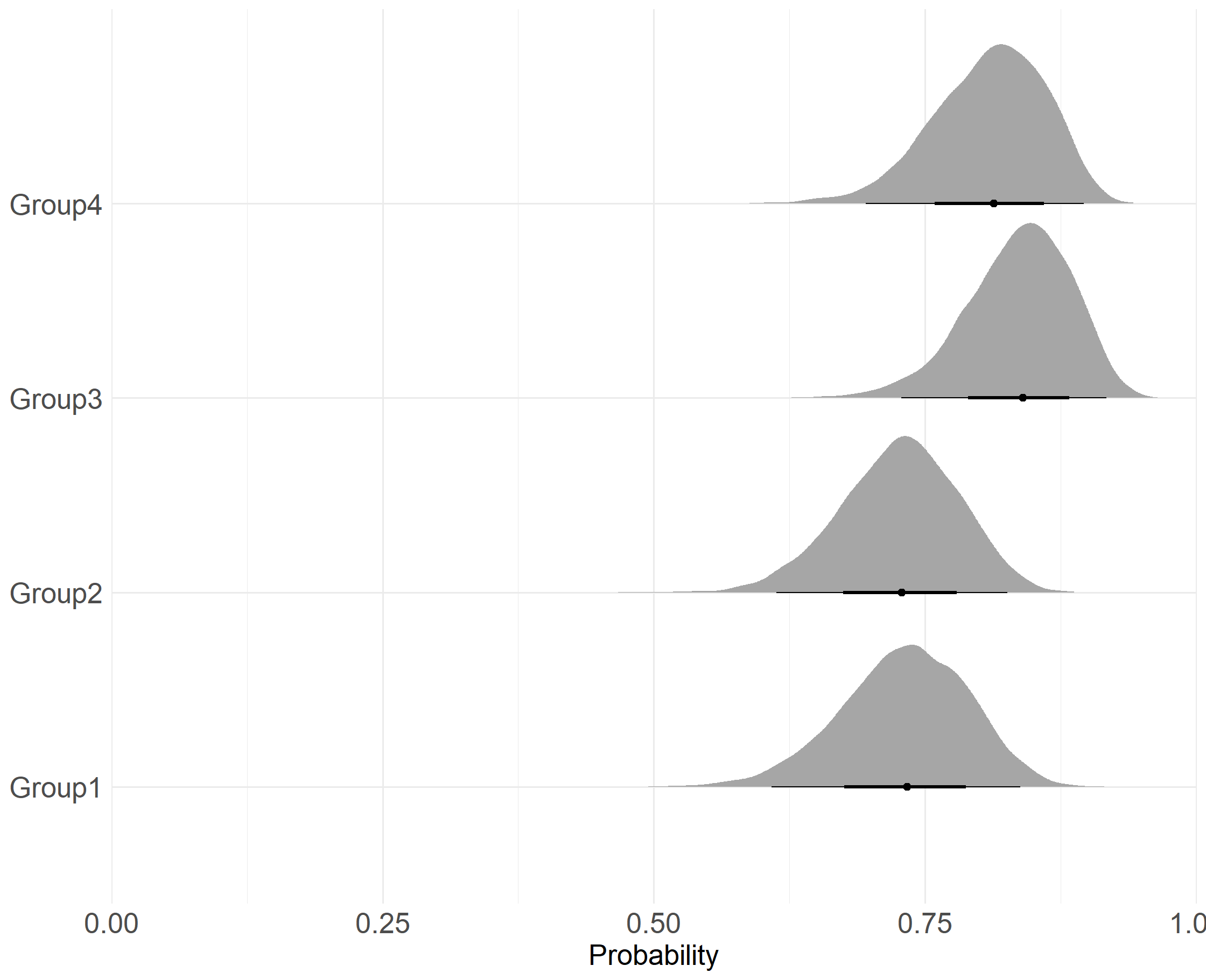, width = 12cm}}
  \caption{Posterior distribution of Group Probabilities}
  \label{fig:GroupPostDistribution}
\end{figure}

The intercept, representing the estimated log-odds of a correct answer for the reference group (Group 4), was 1.48 (95\% CI: 0.84, 2.18), which corresponds to an estimated probability of approximately 0.82 (95\% CI: 0.70, 0.91) (see Figure \ref{fig:GroupPostDistribution}). For Group 1 (pressure + gloves), the estimated log-odds difference versus control was –0.46 (95\% CI: –1.33 to 0.38), corresponding to about a 85\% chance that the condition truly reduces recall. A Bayes factor of 5.83 indicates moderate evidence in favor of a negative effect. Similarly, Group 2 (pressure only), the estimated log-odds difference against group 4 (control) is –0.49 (95\% CI: –1.33 to 0.33), corresponding to about an 88\% probability that the pressure alone reduces recall. A Bayes factor of 7.1 indicates moderate evidence in favor of a negative effect. For group 3 (glove only), the estimated log-odds difference for gloves only versus group 4 (control) indicated a small positive effect (0.21; 95\% CI: –0.69 to 1.14) with very high uncertainty. There was a 33\% probability that this small effect is negative (i.e., that gloves reduce recall), and the Bayes factor of 0.5 slightly favors a non-negative effect. Posterior predictive checks indicate our observed mean (7.8) lies at the center of the model’s simulated means, demonstrating good fit (Bayesian $ p \approx 0.5$).


\subsection{Bayesian Mediation Analysis}

A Bayesian multivariate model was used using
the brms package in R to examine whether perceived workload (rescaled to the [0,1] interval) mediated the effect of pressure on information retention. The model included the two factors: pressure \& gloves, and their interaction as predictors of both perceived workload and information retention. The perceived workload was modeled using a Beta distribution with the following priors: a normal prior of $normal(0.4, 0.3)$ for the intercept, reflecting moderate expected workload; weakly informative slope priors of $normal(0, 0.3)$; and a $Gamma(30, 2)$ prior for the precision parameter (phi), assiming a weak effect of pressure and gloves on information retention and perceived workload. The mental effort (raw reaction time in seconds) was modeled with a lognormal distribution with a $Normal(1.77, 0.3)$ prior on the intercept—reflecting an expected mean reaction time of $\approx$5 seconds—and weakly informative $Normal(0, 0.3)$ priors on the slopes for pressure, gloves, and their interaction. Finally, a half-Normal prior with $normal(0,0.2)$ to encode the modest trial-to-trial variability in log‐RT. For the information retention, the priors from the previous Section \ref{subsecInfoRetention} were used (see Figure \ref{fig:PriorPostDistribution}). 

Prior predictive check was conducted for perceived workload and mental effort, confirming that the priors generated plausible simulated data consistent with our weak informative stance. The multivariate Bayesian model was estimated using four MCMC chains, each with 4000 iterations, including 2000 warm-up iterations, resulting in 8000 post-warmup samples. Convergence was assessed using the potential scale reduction factor $\hat{R}$, which was equal to 1.00 for all model parameters, indicating satisfactory convergence across chains. Furthermore, effective sample sizes (Bulk ESS and Tail ESS) for all parameters exceeded 6000, suggesting high precision and stability in the estimation of both central tendencies and tail distributions of the posterior.

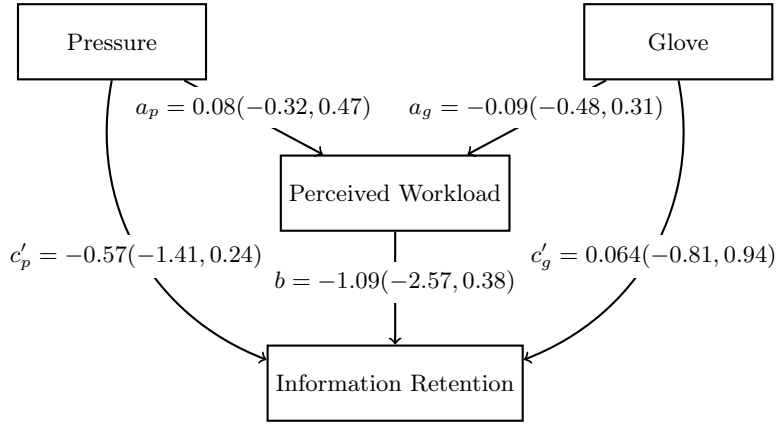
\begin{figure}[!h]
\begin{center}
\begin{tikzpicture}[node distance=1.5cm and 5cm, auto, thick]

\tikzstyle{node} = [draw, rectangle, minimum width=2.5cm, minimum height=1cm]

\node[node] (p) {Pressure};
\node[node, right=of p] (g) {Glove};
\node[node, below=of $(p)!0.5!(g)$] (w) {Perceived Workload};
\node[node, below=of w] (r) {Information Retention};

\draw[->] (p) -- node[above, yshift=-1.5mm,fill=white] {\small $a_{p} = 0.08 (-0.32,     0.47)$} (w);
\draw[->] (g) -- node[above, yshift=-1.5mm,fill=white] {\small $a_{g} = -0.09(-0.48,0.31)$} (w);
\draw[->] (w) -- node[above, yshift=-2mm,fill=white] {\small $b = -1.09(-2.57,0.38)$} (r);
\draw[->] (p) to[bend right=40] node[below, yshift=2mm,fill=white] {\small $c'_{p} = -0.57(-1.41,0.24)$} (r);
\draw[->] (g) to[bend left=40] node[below, yshift=2mm,fill=white] {\small $c'_{g} = 0.064(-0.81,0.94)$} (r);

\end{tikzpicture}
\end{center}
\caption{Mediation diagram for perceived workload}\label{fig:Med_PW}
\end{figure}

The effect of pressure on perceived workload was 0.08 (95\% CI: –0.32, 0.48), indicating a small, uncertain increase in perceived workload under pressure. The credible interval included zero, suggesting a lack of strong evidence for any effect. Similarly, Figure \ref{fig:Med_PW} depicts neither indirect nor direct paths were statistically significant, indicating no evidence of mediation through perceived workload. The path from perceived workload to information retention was negative but uncertain, -1.09 (95\% CI:–2.57, 0.38). Bayesian estimation indicated a 63\% probability that perceived workload negatively mediates the effect of Pressure on Information Retention, and a 65\% probability that it positively mediates the effect of Glove use. However, the 95\% credible intervals include zero in both cases, suggesting limited and uncertain evidence for mediation. Overall, neither intervention showed a clear effect on retention via perceived workload. 

\begin{figure}[!h]
\begin{center}
\begin{tikzpicture}[node distance=1.5cm and 5cm, auto, thick]

\tikzstyle{node} = [draw, rectangle, minimum width=2.5cm, minimum height=1cm]

\node[node] (p) {Pressure};
\node[node, right=of p] (g) {Glove};
\node[node, below=of $(p)!0.5!(g)$] (m) {Mental Effort};
\node[node, below=of w] (r) {Information Retention};

\draw[->] (p) -- node[above, yshift=-1.5mm,fill=white] {\small $a_{p} = -0.12 (-0.44,0.18)$} (w);
\draw[->] (g) -- node[above, yshift=-1.5mm,fill=white] {\small $a_{g} = 0.35(0.01,0.67)$} (w);
\draw[->] (m) -- node[above, yshift=-2mm,fill=white] {\small $b = 0.52(-0.45,1.48)$} (r);
\draw[->] (p) to[bend right=40] node[below, yshift=2mm,fill=white] {\small $c'_{p} = -0.52(-1.35,0.28)$} (r);
\draw[->] (g) to[bend left=40] node[below, yshift=2mm,fill=white] {\small $c'_{g} = -0.08(-1.02,0.86)$} (r);

\end{tikzpicture}
\end{center}
\caption{Mediation diagram for mental effort}\label{fig:Med_ME}
\end{figure}
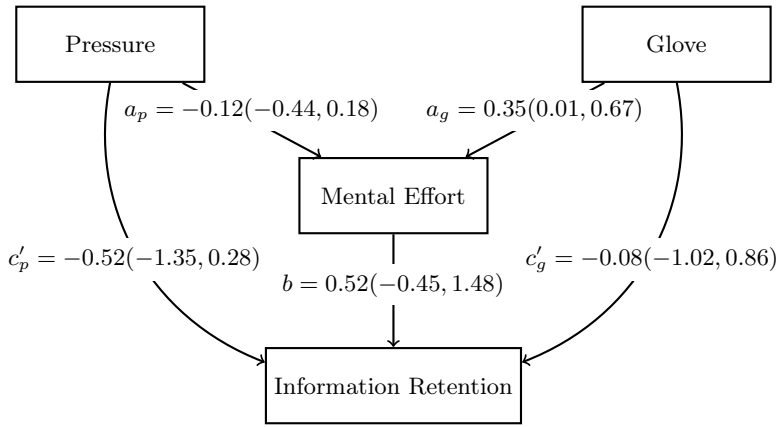

The Pressure condition had a small negative, uncertain effect on mental effort -0.12 (95\% CI: –0.44, 0.18). The credible interval included zero, suggesting a lack of strong evidence for any effect. Although the Glove condition significantly increased reported mental effort, Figure \ref{fig:Med_ME} suggests neither indirect nor direct paths were statistically significant, indicating no evidence of mediation through mental effort. The path from mental effort to information retention was positive but uncertain, 0.52 (95\% CI: –0.45, 1.48). There is a 70\% probability that mental effort negatively mediates the effect of Pressure on Information Retention, and an 85\% probability that it positively mediates the effect of Glove use. In both cases, however, the 95\% credible intervals for the indirect effects include zero, suggesting substantial uncertainty in the mediation effects. Overall, neither intervention showed a clear effect on retention via mental effort.

\section{Discussion}\label{secDiscuss}

In this study, we explored the effect of haptic perception/senses (touch + movement) on information retention through mental effort and perceived workload as mediator variables. Touch sensitivity (as deprivation of tactile perception) was enforced by requiring participants to wear gloves during handwriting (\textbf{RQ1}). Movement (as the intensity of motor activity) was enforced by requiring participants to write with additional pressure (\textbf{RQ2}). The learning performance was measured by the participant's performance on the immediate recall test(10-item test). The mental effort and the perceived workload were recorded as the mediating variables. The mental effort was measured using reaction time in a secondary task using the dual task method. The perceived workload was measured with the Nasa-TLX taskload questionnaire. 

We used a Bayesian binomial regression with a logit link to model information retention, applying weakly informative priors refined through prior predictive checks. The posterior for the control group suggested an 82\% average correct response rate (95\% CI: 70\%–91\%). Compared to control, group 1 with added pressure (movement) + wearing gloves (touch) and Group 2 with added pressure only, both showed moderate evidence of reduced recall, with posterior probabilities of 85\% and 88\% for negative effects and Bayes factors (BF) of 5.83 and 7.1, respectively. Gloves only (Group 3) showed a small, highly uncertain positive effect (33\% probability of being negative; BF = 0.5). Posterior predictive checks indicated good model fit. Overall, there was moderate evidence suggesting that both pressure conditions (with or without gloves) negatively impacted information retention, whereas the glove-only condition showed no clear evidence of an effect. Furthermore, all groups performed relatively well in the information retention test. Since learning in the context of this study was purely cognitive, that is, assimilation and recall from memory, our results are in line with \cite{ray_relationship_2022}, who only found weak evidence for the effect of psychomotor aspects on cognitive learning. The memorization and conceptualization benefits of handwriting \cite{flanigan_typed_2024,wrigley_avoiding_2019,mueller_pen_2014} were assumed to be the result of the cognitive engagement due to the time constraint, which required learners to actively process information to condense it. For example, \cite{richardson_which_2024} found that the drawing condition, which necessitates learners to physically and visually depict each word—a form of contextualized, multimodal activity—resulted in significantly better recall than handwriting or typing. 

Using Bayesian multivariate analysis, we found no clear evidence that perceived workload or mental effort mediated the effect of pressure or gloves on information retention. Although small trends suggested potential negative mediation via workload (63\% probability) and positive mediation via mental effort for gloves (85\% probability), all Credible Intervals included zero, indicating substantial uncertainty and weak evidence for any mediation effects. Mental effort is generally considered to positively impact memory \cite{bjork_making_2011}; \cite{peverly_importance_2006} cautions against the possibility of handwriting to cognitively overload learners. The small potential negative mediation via workload, which measures both mental and physical effort, potentially hints towards physical effort playing a negative role outside the mental effort constraints. However, the physical act of handwriting is contextually coupled with the mental processes involved in processing incoming information, and as most participants have already internalised handwriting, the treatments may not have had the intended effect. This aligns with the remarks of \cite{danna_basic_2015} and potentially hints towards the integration of haptic modality in the Modality principle in Multimedia Learning \cite{Mayer_2005}, which suggests that the use of multiple modalities results in efficient learning when the information is contextually coupled. 

\subsection{Theoretical and Practical Implications}
Sensors are indispensable for the study of embodied learning and complex skills development through authentic practice. Paradoxically, the use of such sensors and actuators, which are added on top of the actual authentic settings, can impact learning. For example, \cite{mat_sanusi_table_2021} used smartphone sensors by attaching a smartphone to the learner's body, arguably affecting the learner's authentic performance. Thus, understanding the effect of such additions on learning is of utmost importance, not only for designing authentic learning environments but also for studying embodied learning theories in situated contexts. 

The results of this study also suggest that wearing a glove and/or exerting additional pressure during handwriting does not affect recall, mental effort, or perceived workload. Such additions in the context of handwriting may not cognitive overload the learner. However, \cite{vermeulen_sensory_2008} found that sensory overload resulted in cognitive overload when additional stimuli were presented, but the increase in the intensity of existing primary stimuli seemed to have no impact. This potential to provide additional feedback and support during authentic practice may improve learning and acquisition of complex skills \cite{danna_basic_2015}.

This may suggest that wearable sensors can potentially be safely used for learning handwriting.  \cite{doug_handwriting_2019} found that the students’ handwriting performance is continuously degrading, affecting their academic performance. While occupational therapy-based interventions have proven beneficial \cite{hoy_systematic_2011}, they cannot address the issue at the required scale. Immersive learning technologies that support authentic practice can contribute towards solving this problem by automating educational aspects surrounding handwriting. For example, automated systems utilizing consumer tablets have been developed to diagnose handwriting difficulties such as dysgraphia \cite{asselborn_extending_2020}.

Similarly, \cite{dikken2022deliberate} developed a sensor-based application for training handwriting that provides real-time feedback on various handwriting attributes based on the teacher's expertise. Such attributes, like pen pressure and perceptual-motor abilities, directly impact handwriting itself \cite{dennis_pencil_2001}, which further impacts academic performance. The potential to make use of (invasive) sensors, such as the SenseGlove\texttrademark, potentially without adversely affecting learning, broadens the horizons for more inclusive multisensory and seamless learning design \cite{specht2019sensors}. 

\subsection{Limitations}
The study is limited by the sample size (N=20). Despite the use of Bayesian analysis, which does not necessarily rely on a large number of participants, the posteriors remain uncertain, reflecting the limited data. 

The study was constrained by the use of a structured text excerpt as the basis for the recall task. This contrasts with prior research employing lists of unrelated words to assess memory performance \cite{bouriga_is_2021,van_der_weel_handwriting_2024}. Structured text excerpts may not necessarily overload the working memory, especially in the presence of prior knowledge. Notably, no prior-test was conducted to assess participants’ baseline knowledge. However, the text was deliberately selected from a geology textbook under the assumption—based on participants’ enrollment in computer science programs—that prior familiarity with the content would be minimal. Using a text excerpt was a conscious choice to test the effects of handwriting in the absence of repetition.

The experimental treatment involved manipulating two modalities of the haptic sense, but each was only altered in one direction, even though opposite manipulations might also influence learning. For example, writing on paper is superior in terms of brain activation in comparison to writing on digital tablets \cite{umejima_paper_2021}, conceivably due to the increased friction provided by the paper's rough texture. In this study, the touch or tactile perception was reduced, and the writing pressure was increased. However, the corresponding reverse manipulations were not done due to logistical constraints. 

Lastly, the NASA Task Load Index (NASA TLX) questionnaire used in the study consists of only 14 categories compared to its 15 categories. One category was removed as it was irrelevant to this study. This discrepancy can impact the reliability of the instrument.

\section{Conclusion}\label{secConclusion}
In this exploratory pilot study, we experimentally investigated the effect of manipulating two attributes associated with haptic perception/sense, namely touch and movement, on learning. Learning is defined as the assimilation and recall of information from memory. Mental effort and perceived workload were observed as the mediating variables. Perceived workload is the combined mental and physical effort, which was included due to the emphasis on the haptic senses. The tactile perception or touch was manipulated by wearing a glove, while the movement was manipulated using software to enforce higher pressure while writing. Moderate evidence suggested that increased writing pressure negatively affected information retention, while glove use showed no clear effect. No evidence was found that perceived workload or mental effort mediated these effects, with all estimates remaining uncertain. The findings of this study may contribute to our understanding of the design of authentic learning environments with sensors and actuators. It may also contribute to future studies aiming to understand embodied theories and how the potentialities of our bodies and environments impact cognitive processes.

\begin{credits}
\subsubsection{\ackname} 
We want to thank the students at the University of Duisburg-Essen who helped execute the study in various roles. Namely, \textit{Anna Luisa Färber}, \textit{Bünyamin Yilmaz}, \textit{Hannah Holland}, \textit{Joshua Jung}, \textit{Katarzyna Pogorzala}, \textit{Van Hoang}, and \textit{Zhang Xinyu}.

\subsubsection{\discintname}
The authors have no competing interests to declare that are relevant to the content of this article
\end{credits}

%
%
%
 \bibliographystyle{splncs04}
 \bibliography{references}
\end{document}